\documentclass[prl,aps,twocolumn,showpacs,10pt]{revtex4-1}
\usepackage[dvips]{graphicx}
\usepackage{latexsym}
\usepackage{amsmath}

\newcommand{\vect}[1]{\boldsymbol{#1}}

\begin{document}


\title{Theory of a continuous stripe melting transition in a two dimensional metal: Possible application to cuprates}
\author{David F. Mross, and T. Senthil}
\affiliation{Department of Physics, Massachusetts Institute of Technology,
Cambridge, MA 02139, USA }

\date{\today}
\begin{abstract}
 We develop a concrete theory of continuous stripe melting quantum phase transitions in two dimensional metals and the associated Fermi surface reconstruction. Such phase transitions are strongly coupled but yet theoretically tractable in situations where the stripe ordering is destroyed by proliferating doubled dislocations of the charge stripe order. The resulting non-Landau quantum critical point (QCP) has strong stripe fluctuations which we show decouple dynamically from the Fermi surface even though static stripe ordering reconstructs the Fermi surface. We discuss connections to various stripe phenomena in the cuprates. We point out several puzzling aspects of old experimental results (Aeppli et al, Science 1997) on singular stripe fluctuations in the cuprates, and provide a possible explanation within our theory. These results may thus have been the first observation of non-Landau quantum criticality in an experiment.

\end{abstract}
\newcommand{\be}{\begin{equation}}
\newcommand{\ee}{\end{equation}}
\maketitle

Over the last 15 years it has become clear that a tendency to charge/spin stripe order is remarkably common in almost all families of underdoped cuprates\cite{review}. 
Recently the idea that 
quantum criticality associated with the onset of stripe order and associated fermi surface reconstruction may be responsible for the non-fermi liquid physics seen around optimal doping in the normal state has become popular\cite{taillefer}. Despite this strong motivation there is very little theoretical understanding of continuous stripe ordering transitions in a metallic environment. In a weakly interacting metal the stripe ordering transition can be formulated by coupling the fluctuating charge or spin stripe order parameter to the Fermi surface. Recent work shows that the stripe fluctuations are strongly coupled to the Fermi surface at low energies and there is no controlled description\cite{abanov,maxnematic} so that the theory is poorly understood. 

For the cuprates these difficulties are perhaps not directly bothersome since the weak coupling description of the stripe fluctuations is in any case unlikely to be the right starting point. Rather as advocated in Ref. \onlinecite{electronlc}, it may be more fruitful to take a strong coupling point of view and regard the phase transition as a quantum melting of stripe order driven by proliferation of topological defects. A theory of continuous quantum melting phase transitions of stripe order in a metal is not currently available and will be provided in this paper. 

On the experimental side surprisingly little is known about the possible presence of quantum critical stripe fluctuations around optimal doping. An important and well known exception is a neutron scattering study of near optimal $La_{2-x}Sr_xCuO_4$ by Aeppli et al\cite{aeppli}. As we discuss below these results paint a rather intriguing picture of the singular stripe fluctuations.


Ref. \onlinecite{aeppli} measured the spin fluctuation spectrum in $La_{1.85}Sr_{0.14}CuO_4$ over a wide range of frequency and temperature near the spin stripe ordering wavevector. The width of the incommensurate peak (which is the inverse correlation length) increases approximately as
$\sqrt{T^2 + \omega^2}$ as expected of a strongly coupled QCP with dynamical critical exponent $z = 1$. We would like to point out that $z = 1$ is rather surprising for a metal like near optimal LSCO. The spin stripe ordering wavevector clearly connects two points of the electronic Fermi surface measured by photoemission. In any metal that has reasonably sharp quasiparticle-like peaks (certainly in a Landau Fermi liquid which the metallic state is of course not but also in a marginal Fermi liquid and other non-fermi liquid models\cite{stripecritlong} which it might be) the stripe fluctuations will be Landau damped. Usually this Landau damping is strongly relevant and leads to a renormalization of $z$ away from 1. Thus the observation of $z = 1$ is significant. It suggests that either the usual Landau damping mechanism is absent in the normal state or that the Landau damping is present but does not affect the quantum critical fluctuations. The latter possibility is a strong hint of a quantum phase transition beyond the Landau-Ginzburg-Wilson paradigm. 

Further evidence for a non-Landau QCP comes from measurements of the height of the peak of the incommensurate spin fluctuations at zero frequency. The imaginary part of the dynamic spin susceptibility satisfies $ \frac{\chi_P"(\omega, T)}{\omega} \sim \frac{1}{T^2}$ at low frequency. Within $z = 1$ scaling, a standard scaling argument shows that this implies an anomalous exponent $\eta = 1$ for the critical spin fluctuations. Such a large value of $\eta$ is uncommon for Landau QCPs but is typical for non-Landau QCPs \cite{dccp,chub,kim,motvish,sandvik,tarun,isakov}. 

The absence of Landau damping effects is consistent with other observations of magnetic excitations in underdoped cuprates\cite{tranquada}. Pertinent to this is whether a pseudogap is present in the ARPES spectra that partially gaps out the Fermi surface. If the hot spots lie in pseudogapped portions of the Fermi surface no Landau damping may be expected. At doping $x = 0.15$ LSCO has a pseudogap in the ARPES spectrum which opens below 150 K\cite{zx}. The neutron data of Ref. \onlinecite{aeppli} extends from 300 K to 35 K and evolves smoothly without noticing the opening of the pseudogap. Thus it appears as though the critical stripe fluctuations are indifferent to the fate of the Fermi surface. 

In this paper we develop a concrete theory of continuous stripe melting quantum phase transitions in a metal and use it to propose an explanation of the puzzles pointed above. For concreteness we consider an orthorhombic crystal (tetragonal symmetry will be analysed elsewhere\cite{stripecritlong}) with uni-directional stripe order at some wavevector {\bf Q}. The spin at site ${\bf r}$ varies as 
\begin{align}
\vec S = e^{i {\bf Q}\cdot {\bf r}}\vec M + c.c.
\end{align}
where $\vec M$ is a complex three component vector. 
This kind of spin-order will induce charge order at $2\vect Q$
\begin{align}
\rho_r \sim e^{2i \vect Q\cdot {\bf r}} \psi + c.c.,
\end{align}
with $\psi \sim \vec M^2$. 
For strong coupling stripe melting transitions it is natural to expect that the spin stripe order will melt through two phase transitions - first the spin order goes away while charge stripe order persists ( {\em i.e} translation symmetry remains broken) followed by a second transition where the charge stripe also melts.
Despite the presence of two distinct quantum phase transitions the somewhat higher-$T$ physics will be controlled by a ``mother" multicritical point where spin and charge stripe order simultaneously melt (see FIG. \ref{fig:sn1}). We postulate that the temperature regime probed in the experiments of Ref. \onlinecite{aeppli} is controlled by such a multicritical stripe melting fixed point. We will provide a theory of the critical point where the charge stripe order melts and the multicritical point where spin and charge stripe order simultaneously melt. 

\begin{figure}
\includegraphics[width=\columnwidth]{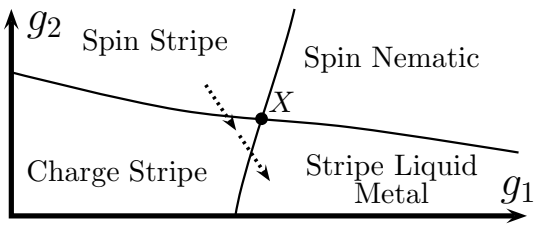}
\caption{Schematic zero-temperature phase diagram close to the multicritical point $X$. The dashed line is parameterized by $g$.}
\label{fig:sn2}
\end{figure}

\begin{figure}
\includegraphics[width=\columnwidth]{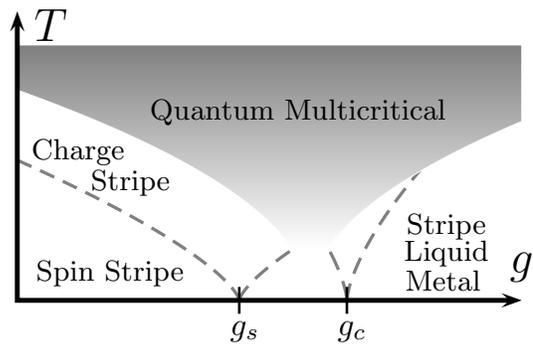}
\caption{Schematic finite temperature phase diagram as a function of $g$ (see FIG. \ref{fig:sn2}). The spin-order vanishes first at $g_s$ while the charge-order persists up to $g_c$.}
\label{fig:sn1}
\end{figure}

\begin{figure}
\includegraphics[width=\columnwidth]{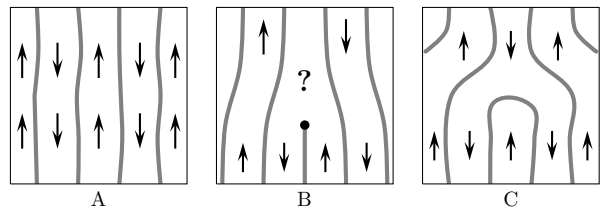}
\caption{$A$: Charge order at $2 \vect Q$ is frequently accompanied by spin-order at $\vect Q$. In this case the spin-order parameter undergoes a sign change from one charge-stripe to the next. $B$: Single dislocations in the charge stripes are bound to half-dislocations for the spin-order, leading to frustration. $C$: Double dislocations in the charge stripes avoid frustration.}
\label{fig:dis}
\end{figure}

Melting of stripe order occurs through proliferation of topological defects. We focus on dislocations in the charge stripe order parameter. To understand the nature of these dislocations we note that the complex vector $\vec M$ may be written 
\begin{equation}
\label{spinstripeop}
\vec M = e^{i\theta_s} \vec N
\end{equation}
with $\vec N$ a real three component vector. The charge stripe order parameter $\psi$ may then be written as 
\begin{equation}
\psi = e^{i\theta_c}
\end{equation}
with $\theta_c = 2\theta_s$. Stripe dislocations correspond to vortices in $\theta_c$. If $\theta_c$ winds by an odd multiple of $2\pi$, then $\theta_s$ winds by an odd multiple of $\pi$. Single valuedness of the spin stripe order parameter implies that $\vec N$ also changes sign on going around such a dislocation. If $\theta_c$ winds by an even multiple of $2\pi$, $\vec N$ is single valued. Thus the spin order is frustrated around odd strength dislocations but not for even strength dislocations (see FIG. \ref{fig:dis}). Formally Eqn. \ref{spinstripeop} contains a $Z_2$ gauge redundancy associated with letting $\vec N \rightarrow - \vec N$, $\theta_s \rightarrow \theta_s + \pi$ at each lattice site. Odd strength dislocations are bound to vortices of the $Z_2$ gauge field. 

Frustration of spin order at a single dislocation contributes a term to its energy that raises it compared to the energy of doubled dislocations\cite{footnote0}. 
This can occur even if there is no long-range spin order but substantial short range spin stripe correlations. If this contribution dominates then it is energetically favorable to proliferate doubled rather than single dislocations. The corresponding stripe liquid phase was first envisaged by Zaanen\cite{zaa} and co-workers and studied further in Refs. \onlinecite{demler,subir,zaanen}. It is strictly distinct (meaning cannot be smoothly connected to) the usual weakly interacting Fermi liquid. However the distinction is extremely subtle and may easily escape detection by any conventional experimental probe. 

When single stripe dislocations have finite core energy, at low energies both $\vec N$ and $b \equiv e^{i\theta_s}$ become well defined. We may envisage four different phases (see FIG. \ref{fig:sn2}). The spin stripe ordered phase has $\langle b \rangle, \langle \vec N \rangle \neq 0$, while a phase with charge stripe but no spin stripe order has $\langle b \rangle \neq 0, \langle \vec N \rangle = 0$. A phase with $\langle b \rangle = 0, \langle \vec N \rangle \neq 0$ preserves translational and time reversal symmetries but breaks spin rotational symmetry by developing a spontaneous spin quadrupole moment $Q_{ab} = N_a N_b - \frac{1}{3}\vec N^2 \delta_{ab}$ (such a phase is also called a spin nematic). Finally $\langle b \rangle = 0, \langle \vec N \rangle = 0$ describes a phase with no broken symmetries but a fractionalization of the stripe order parameter. In the presence of itinerant fermions this latter phase has a conventional large Fermi surface while the striped phases with $\langle b \rangle \neq 0$ will have their Fermi surfaces reconstructed by the stripe order. 

The phase $\theta_c(\vec x, t)$ describes the local displacement of the charge stripes in the $\hat{x}$-direction at time $t$ (we take the stripes to run along $\hat{y}$) and its conjugate variable generates translations of the stripes along $\hat{x}$. An effective model that describes charge stripe fluctuations takes the form of a quantum $XY$ model for the phase $\theta_c$, where the analog of a chemical potential term that couples linearly to the conjugate momentum is prohibited as it is odd under both lattice reflection about $\hat{y}$ and time reversal.

We now discuss the charge stripe melting transition, initially ignoring the coupling of the stripe order parameter to the Fermi surface, and the pinning of the stripe order parameter by the underlying crystalline lattice. 
As the $Z_2$ gauge flux is gapped everywhere in the phase diagram we may safely ignore it to study low energy properties. A `soft-spin' effective field theory that captures the universal properties of all the phase transitions may then be written down in terms of the $b, \vec N$ fields:
\begin{align}
S[b, \vec N] & = \int d\tau d^2x {\cal L}_b + {\cal L}_N + {\cal L}_{bN} \\
{\cal L}_b & = |\nabla b|^2 + \frac{1}{v_c^2}|\partial_\tau b|^2 + r_b |b|^2 + u_b |b|^4 \\
{\cal L}_N & = |\nabla \vec N|^2 + \frac{1}{v_s^2}|\partial_\tau \vec N|^2 + r_N |\vec N|^2 + u_N|\vec N|^4 \\
{\cal L}_{bN} & = v|b|^2 |\vec N|^2
\end{align}
In the spin disordered phases the $\vec N$ field is gapped and may be integrated out. Thus the charge stripe melting 
 is described as an $XY$ condensation transition of the $b$-field. 
However the physical stripe order parameter $\psi = b^2$ is a composite of the fundamental $XY$ field $b$. Thus the stripe order parameter has critical power law correlations with a large anomalous dimension $\eta_\psi \approx 1.49$\cite{sudupe}. This makes the universality class of the transition fundamentally different (but simply derivable) from the ordinary $XY$ universality class. Indeed the only physical operators at this transition are those that are invariant under the local $Z_2$ gauge transformation. For these reasons this transition has been dubbed the $XY^*$ transition in the prior literature \cite{tsfisher,tarun,isakov}.

Including the presence of a metallic Fermi surface leads, in the stripe ordered phase, to a term
\begin{equation}
\label{stripecouple}
g \psi \sum_k c^\dagger_{k+ Q} c_{k} + h.c.
\end{equation}
in the conduction electron Hamiltonian which will reconstruct the Fermi surface if the stripe ordering wavevector connects two points of the Fermi surface. In this case at the critical point or in the stripe melted phase this coupling will lead to the standard Landau damping of the stripe fluctuations:
\begin{equation}
\label{ldcs}
\lambda_d \int d\omega d^2 q |\omega| |\psi (q,\omega)|^2
\end{equation}
The relevance/irrelevance of this term (which is a long ranged imaginary time interaction) at the stripe melting $XY^*$ critical point is readily ascertained by power-counting. Under a renormalization group transformation $x \rightarrow x' = \frac{x}{s}, \tau \rightarrow \tau' = \frac{\tau}{s}$, we have
$\psi \rightarrow \psi' = s^{\Delta} \psi$ with $\Delta = \frac{1+ \eta_\psi}{2}$. This implies
\begin{equation}
\lambda_d' = \lambda_d s^{1 - \eta_\psi}
\end{equation}
As $\eta_\psi > 1$ at the $XY^*$ fixed point, the Landau damping of the critical stripe fluctuations is irrelevant.
The energy density associated with the stripe fluctuations can also couple to the gapless modes of the Fermi surface. As argued in Ref. \onlinecite{subir} these are irrelevant so long as the correlation length exponent $\nu_{XY} > \frac{2}{3}$, which is the case at the $XY^*$ fixed point (in the presence of Coulomb interactions these modes will be suppressed, rendering this coupling even more irrelevant).

Pinning of the stripe order by the underlying lattice is important for commensurate stripes such as the period-4 charge stripe. At the critical point this leads to $8$-fold anisotropy for the $b$ field which is known to be strongly irrelevant at the $2+1$-D XY fixed point. 
 Thus the $XY^*$ stripe melting critical point survives unmodified by either the coupling to the lattice or to the electronic Fermi surface. On tuning through this transition the Fermi surface undergoes a reconstruction, through the coupling of the stripe order parameter to the conduction electrons as described above in Eqn. \ref{stripecouple}. A simple power counting argument\cite{stripecritlong} shows that the gap $\Delta_{FS}$ that opens at the hot spot scales with the distance to the critical point $\delta$ as $\Delta_{FS} \sim |\delta|^{\nu\eta_\psi}$ while the stripe ordering itself occurs at an energy scale $|\delta|^{\nu z}$. As $\eta_\psi>z$ the Fermi surface reconstructs at a scale that is parametrically smaller than the scale of stripe ordering.

Now we turn our attention to the multicritical point where spin and charge stripe order melt simultaneously, i.e. both $b$ and $\vec N$ are critical, ignoring initially both the lattice pinning and the coupling to the Fermi surface. When $v = 0$ the multicritical point where $b$ and $\vec N$ both go critical is described by a decoupled $O(3) \times O(2)$ fixed point where a small $v$ term is an irrelevant perturbation\cite{aharony}, thus there is a finite basin of attraction. 

At this decoupled fixed point the considerations above show that the coupling of the Fermi surface to the charge stripe order prameter $b$ (and lattice pinning for commensurate period-4 stripes) are irrelevant. What about the Fermi surface coupling to $\vec N$? First the coupling of the breathing mode of the Fermi surface to the energy density of $\vec N$ fluctuations is irrelevant as 
$\nu_{O(3)} > \frac{2}{3}$. Second $\vec N$ itself cannot directly couple to the particle/hole continuum at the hot spots of the Fermi surface as it is not gauge invariant. Rather what couples is the physical spin stripe order parameter $\vec M = b\vec N$. The correlations of $\vec M$ in spacetime factorize into a product of the $b$ and $\vec N$ correlators at the decoupled fixed point: 
\begin{equation}
\langle \vec M({\bf x}, \tau)\cdot \vec M( 0, 0)\rangle \sim \frac{1}{\left({\bf x}^2 + v_c^2\tau^2\right)^{\frac{1+\eta_b}{2}}\left({\bf x}^2 + v_s^2\tau^2\right)^{\frac{1+\eta_N}{2}}}
\end{equation}
It follows that the $\vec M$ correlations have anomalous dimension $\eta_M = 1 + \eta_N + \eta_b$ where $\eta_{N,b}$ are the order parameter anomalous dimensions at the $O(3), XY$ fixed points respectively. 
The coupling of $\vec M$ to the Fermi surface particle/hole continuum will generate a Landau damping term
\begin{equation}
\int d^2q d \omega |\omega| |\vec M|^2
\end{equation}
 By the same argument as below Eqn. \ref{ldcs}, this is irrelevant so long as $\eta_M > 1$ which is clearly satisfied. Another gauge invariant operator is the spin quadrupole operator $Q_{ab} = N_a N_b - \frac{1}{3} \vec N^2 \delta_{ab}$ which has scaling dimension $\Delta_Q = \frac{1+\eta_Q}{2}$ with $\eta_Q \approx 1.43$\cite{sudupe}. This has slow correlations near zero wavevector and so couples to the entire Fermi surface. However the coupling to the Fermi surface is only through four fermion terms. Consequently the damping of the $Q_{ab}$ fluctuations by the Fermi surface is weak\cite{stripecritlong} $\sim |\omega|^3|Q|^2$ and is irrelevant. 

We are thus left with the remarkable situation that the decoupled multicritical point survives the inclusion of the coupling to the Fermi surface (and lattice pinning). Clearly finite-$T$ correlations will satisfy $\omega/T$ scaling, with dynamical critical exponent $z = 1$ despite the presence of the particle/hole excitations of the metal. Finally the structure of the critical spin stripe correlations determines the behavior of the dynamical spin susceptibility. A standard scaling argument shows that the temperature dependence of $\frac{\chi_P"(\omega, T)}{\omega}$ measured in the experiments of Ref. \onlinecite{aeppli} goes as $\frac{1}{T^{3 - \eta_M}}$. Using $\eta_{b} \approx 0.04$ and $\eta_{N} \approx 0.04$ we find $\frac{\chi_P"(\omega, T)}{\omega} \sim \frac{1}{T^{1.92}}$ in excellent agreement with the data of Ref. \onlinecite{aeppli}. 

Thus our proposed theory resolves the puzzles posed by the singular stripe fluctuation spectrum. Several concrete predictions also follow from the theory for future experiments. 
First the charge stripe order parameter will exhibit quantum critical scaling with an anomalous dimension $\eta_\psi \approx 1.49$. Second so will the spin quadrupole (i.e spin nematic) order with anomalous dimension $\eta_Q \approx 1.43$. A different qualitatively non-trivial prediction of our theory is the possible existence of the stripe fractionalized metal phase in the overdoped side of the cuprate phase diagram. This phase has a conventional large Fermi surface of electronic quasiparticles and so can be easily mistaken for an ordinary Fermi liquid. The distinction with the Fermi liquid appears in very subtle ways - the soft stripe fluctuations will have a different character from a Fermi liquid, and there will be stable gapped topological defects associated with the remnants of uncondensed single dislocations of the charge stripe order. Thus this phase might have escaped identification in all experiments done to date.

Let us conclude by reiterating our main results. We presented a concrete non-Landau theory of a continuous charge stripe melting transition in a two dimensional metal. The critical stripe fluctuations decouple from the Fermi surface. Despite this, static stripe ordering reconstructs the Fermi surface, though at a scale parametrically different than that of stripe ordering. We discussed various puzzles posed by the existing well known experimental observation\cite{aeppli} of singular spin stripe fluctuations in a near optimal cuprate metal. We proposed an explanation of these puzzles in terms of a non-Landau multicritical quantum stripe melting transition where the spin and charge stripe orders simultaneously melt. Thus Ref. \onlinecite{aeppli} may have been the first experimental observation of non-Landau quantum criticality. We outlined a number of predictions for future experiments. 
For the stripe melting transitions discussed here the decoupling of the critical fluctuations from the Fermi surface means that the Landau quasiparticle is preserved all over the Fermi surface (see Ref. \onlinecite{tarun} for a calculation in a different context). Thus this kind of stripe melting transition cannot explain the observed non-Fermi liquid single particle physics. Since the Fermi surface excitations presumably also determine a host of other non-Fermi liquid properties (such as transport) this kind of stripe melting transition cannot really underlie {\em most} of the observed non-Fermi liquid phenomena. It may however act in parallel with some other mechanism for the destruction of the Landau fermi liquid and can help explain observations related to just the stripe fluctuations and subsequent reconstruction of the Fermi surface.

We thank Eduardo Fradkin, Steve Kivelson, Young Lee and Stephen Hayden for useful discussions. TS was supported by NSF Grant DMR-1005434.

\appendix


\begin{thebibliography}{99}
\bibitem{review} For recent reviews, see e.g. S. A. Kivelson, I. P. Bindloss, E. Fradkin, V. Oganesyan, J. M. Tranquada, A. Kapitulnik, and C. Howald, Rev. Mod. Phys. 75, 1201 (2003); M. Vojta, Adv. Phys. 58, 699 (2009).
\bibitem{taillefer} L. Taillefer, Ann. Rev. Cond. Mat. Phys. 1, 51 (2010).
\bibitem{abanov} A. Abanov and A. V. Chubukov, Phys. Rev. Lett. 84, 5608 (2000); 93, 255702 (2004)
\bibitem{maxnematic} M. A. Metlitski and S. Sachdev, Phys. Rev. B 82, 075127 (2010)
\bibitem{electronlc} S. A. Kivelson, E. Fradkin and V. J. Emery, Nature 393, 550 (1998) 
\bibitem{aeppli} G. Aeppli, T. E. Mason, S. M. Hayden, H. A. Mook, and J. Kulda, Science 278, 1432 (1997)
\bibitem{stripecritlong} David F. Mross and T. Senthil (unpublished). 
\bibitem{dccp}T. Senthil, A. Vishwanath, L. Balents, S. Sachdev, and M. P. A. Fisher, Science 303, 1490 (2004)
\bibitem{chub} A. V. Chubukov, T. Senthil, and S. Sachdev, Phys. Rev. Lett. 72, 2089 (1994).
\bibitem{kim} S. V. Isakov, T. Senthil, and Y. B. Kim, Phys. Rev. B 72, 174417 (2005). 
\bibitem{motvish} O. I. Motrunich and A. Vishwanath, Phys. Rev. B 70, 075104 (2004).
\bibitem{sandvik} A. W. Sandvik, Phys. Rev. Lett. 98, 227202 (2007). 
\bibitem{tarun}T. Grover and T. Senthil, Phys. Rev. B 81, 205102 (2010)
\bibitem{isakov} S. V. Isakov, R. G. Melko, and M. B. Hastings, Science 335, 193 (2012)
\bibitem{tranquada} J. M. Tranquada in \emph{Handbook of High-Temperature Superconductivity}, Springer (2007)
\bibitem{zx} T. Yoshida, M. Hashimoto, S. Ideta, A. Fujimori, K. Tanaka, N. Mannella, Z. Hussain, Z.-X. Shen, M. Kubota, K. Ono, S. Komiya, Y. Ando, H. Eisaki, and S. Uchida, Phys. Rev. Lett. 103, 037004 (2009). 
\bibitem{footnote0} We thank Steve Kivelson for pointing this out to us. Thermal melting of stripes by proliferating double dislocations was studied in Refs. \onlinecite{krueger,bfk,leo}.
\bibitem{krueger} Frank Kr\"uger and Stefan Scheidl, Phys. Rev. Lett. 89, 095701 (2002)
\bibitem{bfk} E. Berg, E. Fradkin, and S. A. Kivelson, Nature Physics 5, 830 (2009).
\bibitem{leo} Leo Radzihovsky and Ashvin Vishwanath, Phys. Rev. Lett. 103, 010404 (2009). 
\bibitem{zaa}J. Zaanen, O. Y. Osman, H. V. Kruis, Z. Nussinov, and J. Tworzydlo , Phil. Mag. B, 81, 1485-1531 (2002)
\bibitem{zaanen}Z. Nussinov, and J. Zaanen, J. Phys. IV 12, 245–250 (2002)
\bibitem{demler}Y. Zhang, E. Demler, and S. Sachdev, Phys. Rev. B 66, 094501 (2002)
\bibitem{subir}S. Sachdev and T. Morinari, Phys. Rev. B 66, 235117 (2002)
\bibitem{sudupe} H. G. Ballesteros, L. A. Fernandez, V. Martin-Mayor, and A. Munoz-Sudupe, Phys. Lett. B 387, 125 (1996) 
\bibitem{tsfisher} T. Senthil and M. P. A. Fisher, Phys. Rev. B 62, 7850 (2000). 
\bibitem{aharony} A. Aharony, Phys. Rev. Lett. 88, 059703 (2002)

\end{thebibliography}
\end{document}